\documentclass[twocolumn,pra,showpacs,amsmath,amssymb]{revtex4}
\usepackage{graphicx}
\usepackage{amssymb,amsthm,amsfonts,amstext}
\usepackage{url}
\def\be{\begin{equation}}
\def\ee{\end{equation}}
\def\bea{\begin{eqnarray}}
\def\eea{\end{eqnarray}}
\def\bma{\begin{mathletters}}
\def\ema{\end{mathletters}}

\def\0{\overline{0}}

\def\q0{\underline{0}}

\def\C{{\mathbb C}}
\def\id{{\mathbb I}}

\def\B{{\cal B}}

\def\tr{\mbox{tr}}
\def\one{\leavevmode\hbox{\small1\normalsize\kern-.33em1}}

\def\bra#1{\langle#1|} \def\ket#1{|#1\rangle}

\def\proj#1{\ket{#1}\!\bra{#1}}

\def\id{{\mathbb I}}
\def\tr{\mbox{tr}}

\begin{document}

\title{Activation of Non-Local Quantum Resources}

\author{Miguel Navascu\'es$^1$ and
        Tam\'as V\'ertesi$^2$}
\affiliation{$^1$Facultad de Matem\'aticas, Universidad Complutense de Madrid\\
$^2$Institute of Nuclear Research of the Hungarian Academy of Sciences}

\begin{abstract}
We find two two-qubit bipartite states $\rho_1,\rho_2$ such that arbitrarily many copies of one or the other cannot exhibit non-local correlations in a two settings/two outcomes Bell scenario. However, the bipartite state $\rho_1\otimes\rho_2$ violates the Clauser-Horne-Shimony-Holt (CHSH) Bell inequality \cite{CHSH} by an amount of 2.023. We also identify a CHSH-local state $\rho$ such that $\rho^{\otimes 2}$ is CHSH-violating. The tools employed can be easily adapted to find instances of non-locality activation in arbitrary Bell scenarios.

\end{abstract}

\maketitle

In 1964 John S. Bell refuted the Einstein-Podolski-Rosen (EPR) argument against Quantum Mechanics \cite{EPR} by showing that certain correlations generated by distant observers measuring a quantum system are impossible to reproduce in any local realistic theory \cite{bell}. Such correlations, which do not arise as marginals of a greater probability distribution, have been termed `non-local' in recent years. Initially a foundational concept, non-locality has lately found applications in several areas of Quantum Information, like quantum cryptography \cite{crypto}, certified randomness generation \cite{random} and device-independent quantum state estimation \cite{state1,state2}. In any concrete implementation of the former protocols, non-locality is quantified according to some linear figure of merit, the so-called Bell functional. The amount by which the Bell value of a given quantum set of correlations differs from the classical maximum (the Bell inequality violation) shall thus be regarded as a resource in these primitives.

A prerequisite for the emergence of non-locality is the presence of entanglement in the quantum state where the measurements are carried out. However, the opposite implication is far from clear. Although there exist examples of entangled quantum systems unable to produce non-local correlations \cite{local}, Popescu and Gisin showed that some of those states exhibit a non-local behavior if, prior to the Bell tests, they are subject to local measurements and post-selected according to the measurement results \cite{popescu,gisin}. Some time later, Peres observed that an even bigger set of local states reveal a non-local nature when post-selections and measurements are conducted, not over a single local state, but over several identical copies \cite{peres}.

In view of Peres' results, it is reasonable to ask whether this form of `non-locality activation' could also be achieved by performing measurements over ensembles of local states only, without resorting to any form of post-selection. This corresponds to a natural situation where the distant parties receive a number of local states and are asked to violate a Bell inequality with no prior communication. Although there have been some numerical attempts to attack this problem \cite{liang}, it is still an open question whether such a `tensoring-mediated activation' exists.

If our figure of merit is not mere deviation from locality, but the violation of a specific Bell inequality, the corresponding scenario is quite similar to the one above: given a Bell functional $\B$, we can define $\B$-local states as those quantum systems where the evaluation of $\B$ cannot exceed the classical bound. Analogously, given a set of $\B$-local states, we may speak of $\B$-activation if those states can be locally engineered to produce a $\B$ violation.

When $\B$ happens to be the Clauser-Horne-Shimony-Holt (CHSH) inequality \cite{CHSH} and local filtering is allowed, the connection between entanglement and non-locality is well understood. We know that the set of quantum states which violate CHSH under local post-selection of many copies coincides with the set of distillable states \cite{masanes1}. Moreover, we know that for any entangled state $\sigma_1$ there exists another state $\sigma_2$ none of whose local filterings violates the CHSH inequality, but such that $\sigma_1\otimes\sigma_2$ violates CHSH after some local post-selection \cite{masanes2}. Choosing $\sigma_1$ undistillable (and so unable to violate CHSH) in the above proposition, we thus have that CHSH non-locality can be activated by post-selecting tensor products of CHSH-local states.

Nothing is known, though, about activation of bipartite Bell inequalities by tensoring alone (note, however, the recent work \cite{cavalcanti}, where the authors show instances of non-locality activation in the multipartite setting). This led Liang to ask in \cite{liang,open} whether there exist two CHSH-local states $\rho_1,\rho_2$ such that $\rho_1\otimes\rho_2$ generates CHSH-violating statistics when measured in the appropriate bases.

In this article we will answer this question in two different ways: first, we will show the existence of two states $\rho_1,\rho_2$ such that $\rho_1^{\otimes N}$ and $\rho_2^{\otimes N}$ do not violate CHSH for any $N$, but $\rho_1\otimes\rho_2$ does. Second, we will construct a CHSH-local state $\rho$ such that $\rho^{\otimes 2}$ is CHSH-violating.

\vspace{10pt}
\noindent\textbf{Some examples of non-locality activation}

Consider a situation where two space-like separated observers, say, Alice and Bob, are conducting measurements on a joint quantum system. Whenever Alice applies an interaction $x\in \{1,...,s_A\}$ on her subsystem, she will read an outcome $a\in \{1,...,d_A\}$ in her detector; likewise, if Bob interacts with his subsystem in some way $y\in \{1,...,s_B\}$, his measurement devices will register a value $b\in \{1,...,d_B\}$. Varying the choice of measurement settings $x,y$, and after many repetitions of the experiment, Alice and Bob would be able to estimate the set of correlations $\{P(a,b|x,y)\}$, where $P(a,b|x,y)$ denotes the probability that Alice and Bob observe the results $a$ and $b$ when they respectively perform the interactions $x$ and $y$. Given such a \emph{$s_As_Bd_Ad_B$ Bell scenario}, we will say that $\{P(a,b|x,y)\}$ is local iff there exists a global measure ${\cal P}(a_1,...,a_{s_A},b_1,...,b_{s_B})$ such that 

\be
P(a,b|x,y)={\cal P}(a_x=a,b_y=b).
\ee

The 2222 Bell scenario is the simplest bipartite setting where one can expect to find locality
violations. There, the locality of a bipartite distribution
$\{P(a,b|x,y):x,y,a,b=1,2\}$ can be decided by checking the value
of the Clauser-Horne-Shimony-Holt (CHSH) inequality \cite{CHSH}.
More concretely, modulo permutations of the settings, $P(a,b|x,y)$ is local iff

\be
|\langle M^1N^1\rangle+\langle M^1N^2\rangle+\langle M^2N^1\rangle-\langle M^2N^2\rangle|\leq 2,
\ee

\noindent where $M^x$ ($N^y$) is any observable that assigns values $+1$ or $-1$ to the two possible outcomes of Alice's (Bob's) measurement $x$ ($y$). Consequently, any bipartite state $\rho$ is local in the 2222 scenario iff, for any set of operators $\{M^1,M^2,N^1,N^2\}$ with spectrum $\{-1,1\}$, the Bell operator

\be
{\cal B}\equiv M^1\otimes N^1+M^2\otimes N^1+M^1\otimes N^2-M^2\otimes N^2
\ee

\noindent is such that $|\tr({\cal B}\rho)|\leq 2$ \footnote{Since any two-outcome measurement is a convex combination of von Neumann measurements, w.l.o.g. we can assume the measurement operators to be projectors.}.

The bipartite quantum states $\rho_1,\rho_2$ alluded to in the abstract of the paper, as
well as the rest of the states and measurements referred to along
this article, can be found in \cite{web}. It can be checked that
$\rho_1^{AB}=\tr_{B'}(\rho_1^{ABB'})$,
$\rho_2^{AB}=\tr_{A'}(\rho_2^{AA'B})$, for some tripartite states
$\rho_1^{ABB'}$ and $\rho_2^{AA'B}$. These states in turn satisfy
$\rho_1^{ABB'}=\rho_1^{AB'B}$ and $\rho_2^{AA'B}=\rho_2^{A'AB}$,
i.e., $\rho_1^{ABB'}$ and $\rho_2^{AA'B}$ are permutationally
invariant with respect to the systems $B,B'$ and $A,A'$,
respectively. Using modern terminology, we would say that $\rho_1$
($\rho_2$) admits a 2-symmetric extension with respect to system
$B$ ($A$) \cite{doherty}. In the following, we will denote by
$S_C^{2}$ the set of all quantum states admitting a 2-extension
with respect to system $C$.

As observed in \cite{terhal}, any state $\sigma^{AB}\in B(H\otimes
H)$ that admits a $2$ symmetric extension (say, with respect to
system $B$) cannot violate any Bell inequality involving $2$
measurement settings on Bob's side. Indeed, denote Bob's
measurements by $\{y_1,y_2\}$, and let
$\{E^a_x,F^{y_1}_{b_1},F^{y_2}_{b_2}\}$ be complete systems of
Positive Operator Valued Measure (POVM) elements, with $x=1,...,s$. Then, one can check that the set
of bipartite correlations $p(a,b|x,y)\equiv\tr(E^a_x\otimes
F^{y}_b\sigma^{AB})$ derives from the global probability
distribution

\be
P(a_1,a_2,...,a_s,b_1,b_2)\equiv p(b_1,b_2)\prod_{x=1}^sp(a_x|x,b_1,b_2),
\label{LHVM}
\ee

\noindent where $p(b_1,b_2)=\tr(\sigma^{BB'}F^{y_1}_{b_1}\otimes F^{y_2}_{b_2})$ and

\be P(a|x,b_1,b_2)\equiv\frac{\tr(\sigma^{ABB'}E^x_a\otimes
F^{y_1}_{b_1}\otimes F^{y_2}_{b_2})}{p(b_1,b_2)}. \ee

\begin{figure}
  \centering
  \includegraphics[width=8.5 cm]{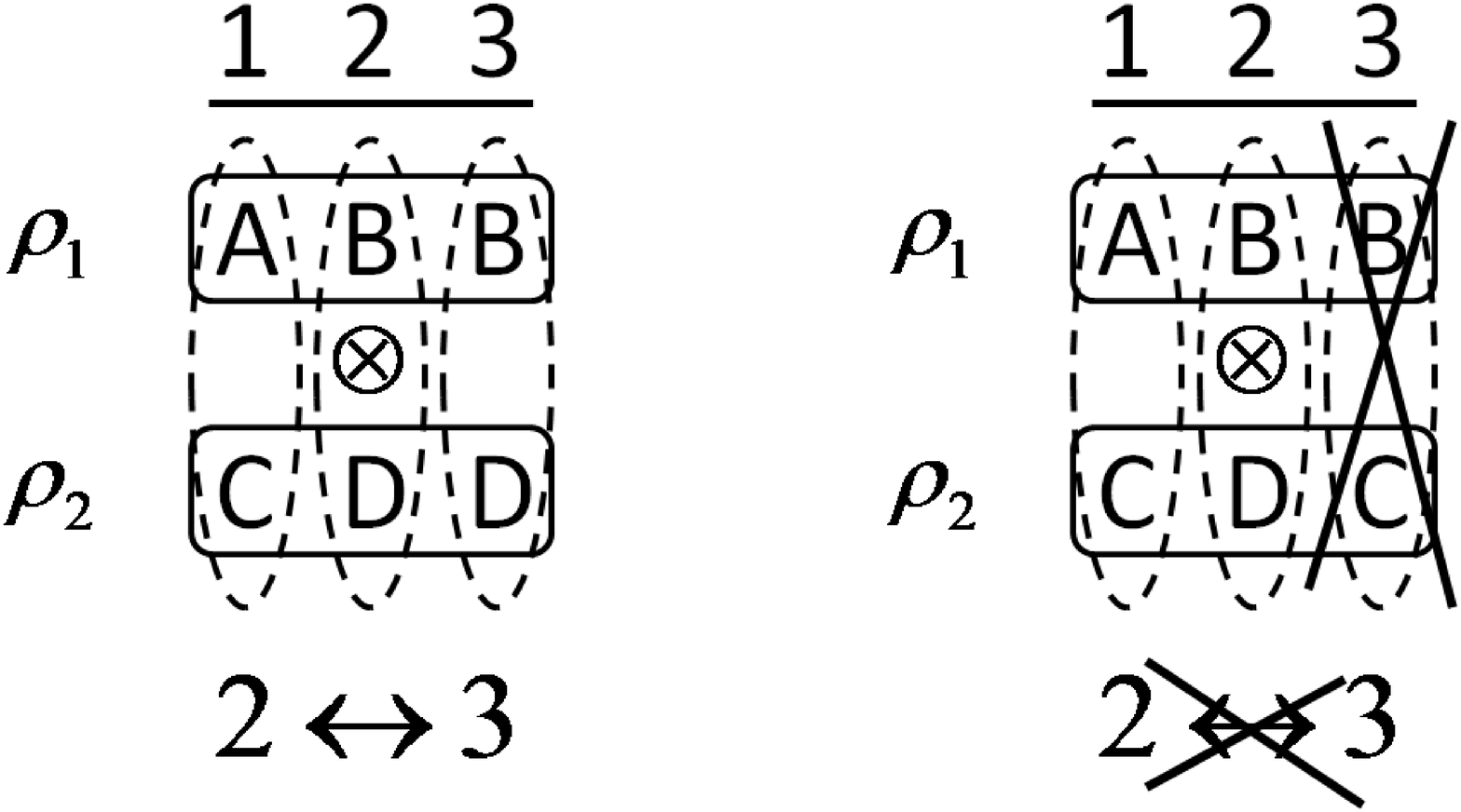}
  \caption{Left: 2-extendable states with respect to a given system (say, 2) are invariant under tensor products; this is so because the tensor product of their extensions is invariant under the interchange of the systems 2 and 3. Right: tensor product of two states admitting 2 symmetric extensions to different systems. In general, these composite states do not admit a 2 symmetric extension.}
  \label{activ_asym}
\end{figure}

As sketched in Figure \ref{activ_asym} (left), if we tensor two
states $\rho_1,\rho_2\in S_B^2$ $(S_A^2)$, the resulting state
$\rho_1\otimes\rho_2$ can also be extended symmetrically with
respect to system $B$ ($A$). It follows that arbitrarily many
copies of $\rho_1$ ($\rho_2$) can never violate any Bell
inequality involving two settings on Bob's (Alice's)
side.

What happens when we join $\rho_1$ and $\rho_2$? Following Figure
\ref{activ_asym}, if $\rho_1\in S_B^2$  and $\rho_2\in S_A^2$,
then the state $\rho_1\otimes\rho_2$ does not necessarily admit a
2-symmetric extension, and so (in principle) could violate CHSH.
And, actually, one can check that the measurement operators given
in \cite{web}, applied to $\rho^{AB}_1\otimes\rho^{A'B'}_2$, result
in a CHSH violation of 2.02324.

In order to come up with the previous example, we used the fact that, for
fixed $(M^x, N^y,\rho_1)$, finding a 2-extendible state $\rho_2$ that maximizes the CHSH value of the system ($M^x,N^y,\rho_1,\rho_2$) is a trivial exercise of linear algebra. In effect, define $\B=M^1_{AA'}\otimes
N^1_{BB'}+M^1_{AA'}\otimes N^2_{BB'}+M^1_{AA'}\otimes
N^2_{BB'}-M^2_{AA'}\otimes N^2_{BB'}$ and
$\B_2\equiv\tr_{AB}(\rho_1^{AB}\otimes \id_{A'B'}\B)$, and let $V_{GH}$ denote the
operator that swaps systems $G$ and $H$. Then,

\begin{eqnarray}
&&\max\{\tr(\B\rho_1\otimes\rho_2):\rho^{A'B'}_2\in S_{A'}^2\}=\nonumber\\
&&=\max\{\tr(\B_2\rho_2):\rho^{A'B'}_2\in S_{A'}^2\}=\nonumber\\
&&=\max\{\tr(\tilde{\B}_2\rho^{CA'B'}):\rho^{CA'B'}\},
\end{eqnarray}

\noindent where $\rho^{CA'B'}$ is an arbitrary normalized quantum
state and $\tilde{\B}_2\in B(H_C\otimes H_{A'}\otimes H_{B'})$ is
given by

\be
\tilde{\B}_2\equiv\frac{1}{2}\{\id_C\otimes \B_2+V_{CA'}(\id_C\otimes \B_2)V_{CA'}\}.
\ee

\noindent The problem just reduces to finding the greatest
eigenvalue of $\tilde{\B}_2$. The corresponding eigenvector
$\ket{\psi}$ can then be transformed into a 2-extendable state
$\rho_2$ through
$\rho_2=\tr_C(V_{CA'}\proj{\psi}V_{CA'}+\proj{\psi})/2$. Likewise,
for fixed $M^x,M^y,\rho_2$, the optimal $\rho_1$ can be found
efficiently.

Conversely, if $(M^x,\rho_1,\rho_2)$ are fixed, one can find the
observables on Bob's side that maximize the CHSH value using the procedure sketched in
\cite{I3322}. That is, given

\begin{eqnarray}
&&F^1\equiv\tr_{AA'}\{([M^1+M^2]\otimes \id_{BB'})(\rho_1\otimes\rho_2)\},\nonumber\\
&&F^2\equiv\tr_{AA'}\{([M^1-M^2]\otimes \id_{BB'})(\rho_1\otimes\rho_2)\},
\end{eqnarray}

\noindent the problem consists in maximizing
$\tr\{N^1F^1+N^2F^2\}$ over all operators $-\id_{BB'}\leq
N^k\leq\id_{BB'}$, $k=1,2$. Now, let $\sum_j
\lambda_j^k\proj{\phi^k_j}$ be the singular value decomposition of
$F^k$. Then, the optimal $N^1,N^2$ are $N^k=\sum_j
\mbox{sgn}(\lambda_j^k)\proj{\phi^k_j}$. Analogously, we can
easily obtain the optimal $M^1,M^2$ for fixed
$N^1,N^2,\rho_1,\rho_2$.

Starting with random states and operators, one can then optimize
one set of variables at a time, following a cycle like
$(M^1,M^2)\to\rho_1\to (N^1,N^2)\to\rho_2\to (M^1,M^2)\to...$.
This optimization procedure frequently gets stuck in local maxima,
so it has to be repeated several times before reaching global
optimality.

We repeated the whole scheme with pairs of states in dimensions
$d=3,4,5$, but we were not able to overcome the activation value
2.040167, which is attained by two pairs of qutrits.

Notice that the former tools can be used to find instances of
activation of any two-setting Bell inequality. If the number of
outcomes of such inequality is greater than two, though, it
becomes necessary to use semidefinite programming (SDP) \cite{sdp}
in order to perform the corresponding POVM optimizations. The
resulting programs are therefore much slower, and wide searches
over high dimensions become computationally expensive.
Nevertheless, we managed to find two states of dimension $3\otimes
3$ that, although unable to violate the three-outcome CGLMP
inequality \cite{CGLMP} separately, could give rise to a CGLMP
value of 2.030126 when taken together.

\begin{figure}
  \centering
  \includegraphics[width=8.5 cm]{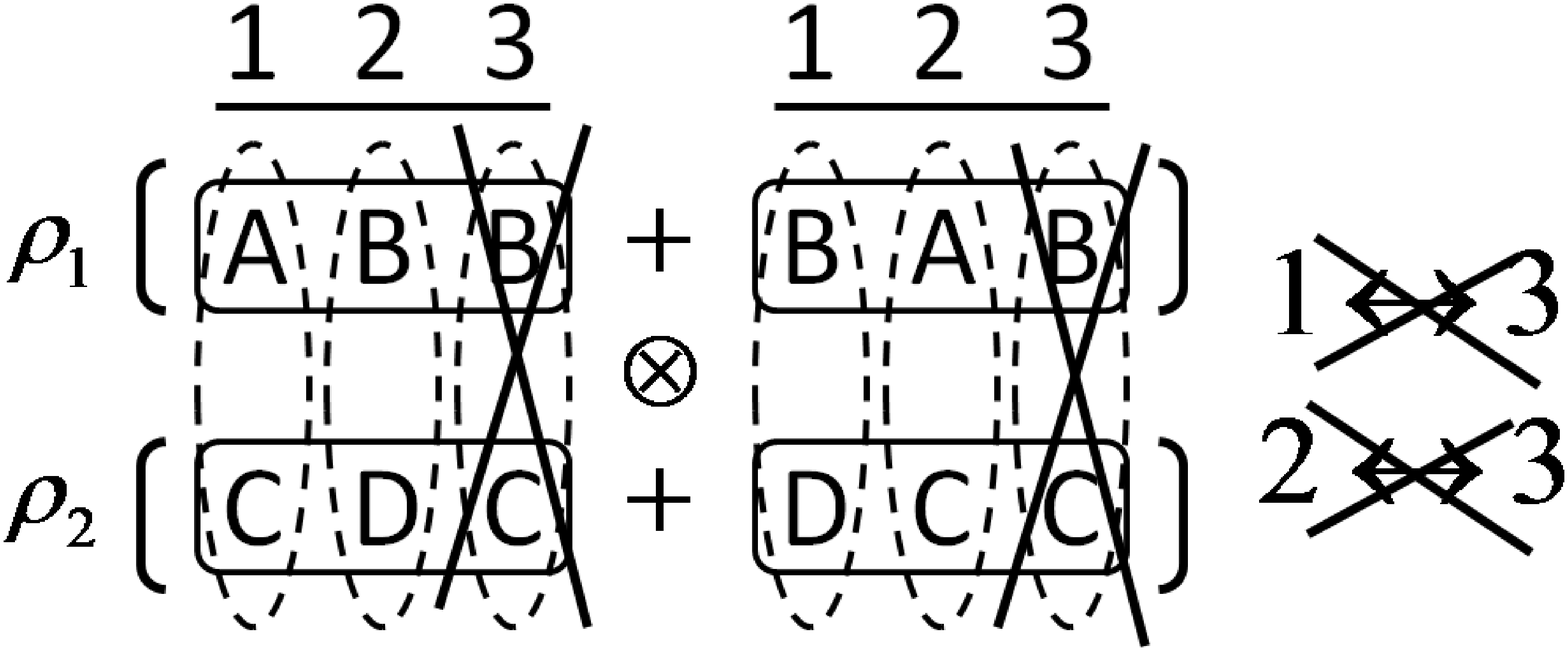}
  \caption{Two bipartite states resulting from the symmetrization of states in $S_B^2$ (or, equivalently, $S_A^2$) are not necessarily local in two-setting scenarios due to the presence of mixed terms admitting different types of extensions.}
  \label{activ_sym}
\end{figure}

\vspace{10pt}
\noindent \textbf{Symmetries}

All states found by our method are highly asymmetric by
construction. Suppose, on the contrary, that we are interested in
finding symmetric states $\rho_1,\rho_2$ with the property that
$\rho_1,\rho_2$ do not violate two-setting Bell inequalities, but
such that $\rho_1\otimes\rho_2$ violates a certain two-setting
Bell inequality. Then one could try the scheme presented in Figure
\ref{activ_sym}, i.e., both $\rho_1$ and $\rho_2$ are the result
of symmetrizing bipartite states $\rho^{AB}_1,\rho^{AB}_2\in
S_B^2$. Being a mixture of $2n$-local and $n2$-local states, such
states are $22$-local.

Like before, it is not difficult to reduce the problem of
maximizing $\tr(M\rho)$ over all possible such states to a
diagonalization problem. Using this new procedure, we were not able to find instances of symmetric activation in dimensions 2 and 3. However, we managed to
find two 22-local symmetric states $\rho_1',\rho_2'\in B(\C^4\otimes \C^4)$
with the property that $\rho'_1\otimes \rho_2'$ violates CHSH by
2.01159. Note, though, that this time locality does not extend to
multiple copies of the states. That is, in principle
$\rho_1^{\otimes N}$ could be CHSH-violating for some $N$. 

This new bound, however, can be beaten easily: take the previous two-qubit asymmetric states $\rho_1,\rho_2$ that produce a CHSH violation of $2.02324$ when taken together, and let $(M^x,N^y)$ be the necessary measurements to ensure so. If, in addition to these systems, Alice and Bob have access to two extra ancillary qubits, then we can take the new symmetric states to be

\begin{eqnarray}
\tilde{\rho}_{1,2}=&&\frac{1}{2}\left(\right.\proj{0}_A\otimes\proj{1}_B\otimes\rho_{1,2}+\nonumber\\
&&+\proj{1}_A\otimes\proj{0}_B\otimes V_{AB}\rho_{1,2} V_{AB}\left.\right).
\end{eqnarray}

\noindent Clearly, these states cannot violate the CHSH inequality. On the other hand, it is easy to check that Alice and Bob can violate CHSH if each of them measures its part of the state $\tilde{\rho}_1\otimes \tilde{\rho}_2$ according to the following measurement scheme: let $z\in\{1,2\}$ be the label of the measurement to be implemented. First, measure the ancillary qubits in the computational basis. If both qubits are in state $\ket{0}$, measure $M^z$ in the rest of the system and output the result (+1 or -1); if the ancillary qubits are both in state $\ket{1}$, measure $N^z$ and output the result; if they are in different states, output 1.

\noindent With this procedure, Alice and Bob should observe a CHSH violation of $(2\times 2.02324+2\times 2)/4 = 2.01162$.

More generally, from any pair of CHSH-local states $\sigma_1,\sigma_2\in B(\C^d\otimes \C^d)$ such that $\sigma_1\otimes\sigma_2$ violates CHSH by an amount $2+\Delta$, one can derive two CHSH-local \emph{symmetric} states $\tilde{\sigma}_1,\tilde{\sigma}_2\in B(\C^{2d}\otimes \C^{2d})$ violating CHSH by $2+\Delta/2$.

This scheme of using simple instances of non-locality activation to construct more sophisticated ones can be carried further. Suppose that we are given any such pair of states $\sigma_1,\sigma_2$. Then one can also find a state $\tilde{\sigma}\in B(\C^{2d}\otimes\C^{2d})$ such that $\tilde{\sigma}$ does not violate CHSH, but $\tilde{\sigma}^{\otimes 2}$ exhibits a CHSH violation of $2+\Delta/2$. Indeed, take the state to be

\be
\tilde{\sigma}\equiv \frac{1}{2}(\proj{1}_A\otimes\proj{1}_B\otimes\sigma_1+\proj{0}_A\otimes\proj{0}_B\otimes\sigma_2),
\label{self_acti}
\ee

\noindent and, as before, let Alice (Bob) measure her (his) main subsystem with $M^x$, $V_{AA'}M^xV_{AA'}$ or $\id_{AA'}$ ($N^y$, $V_{BB'}N^yV_{BB'}$ or $\id_{BB'}$) depending on the state of her (his) ancillary qubits.

Combining both constructions, it is clear that, starting from any CHSH-activating pair $\sigma_1,\sigma_2\in B(\C^d\otimes \C^d)$, we can build a CHSH-local symmetric state $\hat{\sigma}\in B(\C^{4d}\otimes\C^{4d})$ such that $\hat{\sigma}^{\otimes 2}$ violates CHSH by an amount $2+\Delta/4$.

The above results can be easily generalized to arbitrary symmetric Bell inequalities.

%if $\sigma_1,\sigma_2\in B(\C^d\otimes\C^d)$ are $\B$-local but together can produce a violation of $L+\Delta$ (where $L$ is the maximum local value of $\B$), then there exist $\B$-local symmetric states $\tilde{\sigma}_1,\tilde{\sigma}_2\in B(\C^{2d}\otimes\C^{2d}),\hat{\sigma}\in B(\C^{4d}\otimes\C^{4d})$ and a (in general, asymmetric) state $\tilde{\sigma}\in B(\C^{2d}\otimes\C^{2d})$ such that $\tilde{\sigma}_1\otimes\tilde{\sigma}_2,\tilde{\sigma}^{\otimes 2}$ and $\hat{\sigma}^{\otimes 2}$ violate $\B$ by amounts $L+\Delta/2$, $L+\Delta/2$ and $L+\Delta/4$, respectively.

From the previous section, it thus follows that there exists a 22-local symmetric state $\rho\in B(\C^8\otimes\C^8)$ ($\rho'\in B(\C^{12}\otimes\C^{12})$) with the property that two copies of it can violate the CHSH inequality by an amount of 2.00581 (2.01). Likewise, there exists a CGLMP-local symmetric state $\rho''\in B(\C^{12}\otimes\C^{12})$ such that $(\rho'')^{\otimes 2}$ violates the three-outcome CGLMP inequality by 2.007531.

\vspace{10pt}

\noindent \textbf{Conclusion}

In this communication we have given a systematic procedure to
construct examples of non-locality activation. By exploiting the
concept of $2$-extendibility, we were able to find two two-qubit
bipartite states $\rho_1,\rho_2$ such that even though neither
$\rho_1^{\otimes N}$ nor $\rho_2^{\otimes N}$ violates CHSH,
$\rho_1\otimes\rho_2$ produces a CHSH parameter of 2.023. We have
thus solved problem \#21 of the Hannover List of Open Problems in
Quantum Information \cite{open}. 

We have explained how the previous results can be used to find examples of CHSH self-activation, i.e., CHSH-local states $\tilde{\sigma}$ such that $\tilde{\sigma}^{\otimes 2}$ is CHSH-violating. Finally, we have also shown how to extend
our approach to other 2-setting Bell inequalities, like the CGLMP family. Clearly, there
is no need to stop here: our techniques can be easily adapted to
find examples of activation of non-locality in scenarios with more
than two settings per site and more than two sites.

It would be interesting to find out if the small
deviations of locality found here are close to optimal, or, on
the contrary, non-locality activation can be as huge as we
want provided that we go to dimensions high enough. Perhaps the fact that the maximum activation values for CHSH and CGLMP seem to be the same could give us some clue here. We also wonder whether an extreme extension of our results could be given, i.e., whether one could find states $\rho_1,\rho_2$ such that
$\rho_1\otimes\rho_2$ is non-local, but neither $\rho^{\otimes N}_1$ nor $\rho^{\otimes N}_2$ violate \emph{any} Bell inequality.

\noindent\textbf{Acknowledgements}

M. N. has been supported by the European project QUEVADIS. T.V. has
been supported by a J\'anos Bolyai Grant of the Hungarian Academy
of Sciences.

\end{document}